\documentclass[]{aastex631}

\usepackage{epsfig}
\usepackage{amsmath}
\usepackage{amssymb}

\newcommand{\Msolar}{\mbox{$\rm M_{\odot}$}} 

\newcommand{\MMS}{M_{\rm{MS}}}

\shorttitle{BLAPs}
\shortauthors{Zhang et al.}
\graphicspath{{./}{figures/}}

\begin{document}

\title{The formation of blue large-amplitude pulsators from white-dwarf main-sequence star mergers}


\author[0000-0002-3672-2166]{Xianfei Zhang}
\affiliation{Institute for Frontier in Astronomy and Astrophysics, Beijing Normal University, Beijing, 102206, China}
\affiliation{Department of Astronomy, Beijing Normal University, Beijing, 100875, China}
\email{zxf@bnu.edu.cn}

\author[0000-0003-1759-0302]{C.~Simon Jeffery}
\affiliation{Armagh Observatory and Planetarium, College Hill, Armagh BT61 9DG, UK}

\author[0000-0001-7566-9436]{Jie Su}
\affiliation{ Yunnan Observatories, Chinese Academy of Sciences, Kunming 650216, China}
\affiliation{Key Laboratory for the Structure and Evolution of Celestial Objects, Chinese Academy of Sciences, Kunming 650216, China}
\affiliation{International Centre of Supernovae, Yunnan Key Laboratory, Kunming 650216, China}

\author[0000-0002-7642-7583]{Shaolan Bi}
\affiliation{Institute for Frontier in Astronomy and Astrophysics, Beijing Normal University, Beijing, 102206, China}
\affiliation{Department of Astronomy, Beijing Normal University, Beijing, 100875, China}

\begin{abstract}
Blue large-amplitude pulsators (BLAPs) are hot low-mass stars  which show large-amplitude light variations likely due to radial oscillations driven by iron-group opacities.
Period changes provide evidence of both secular contraction and expansion amongst the class.
Various formation histories have been proposed, but none are completely satisfactory.
\citet{Zhang2017} proposed that the merger of a helium core white dwarf with a low-mass main-sequence star (HeWD+MS) can lead to the formation of some classes of hot subdwarf.
We have analyzed these HeWD+MS merger models in more detail.
Between helium-shell ignition and full helium-core burning, the models pass through the volume of luminosity -- gravity-- temperature space occupied by BLAPs.
Periods of expansion and contraction associated with helium-shell flashes can account for the observed rates of period change.
We argue that the HeWD+MS merger model provides at least one BLAP formation channel.
\end{abstract}


\section{Introduction}

\begin{figure}
\plotone{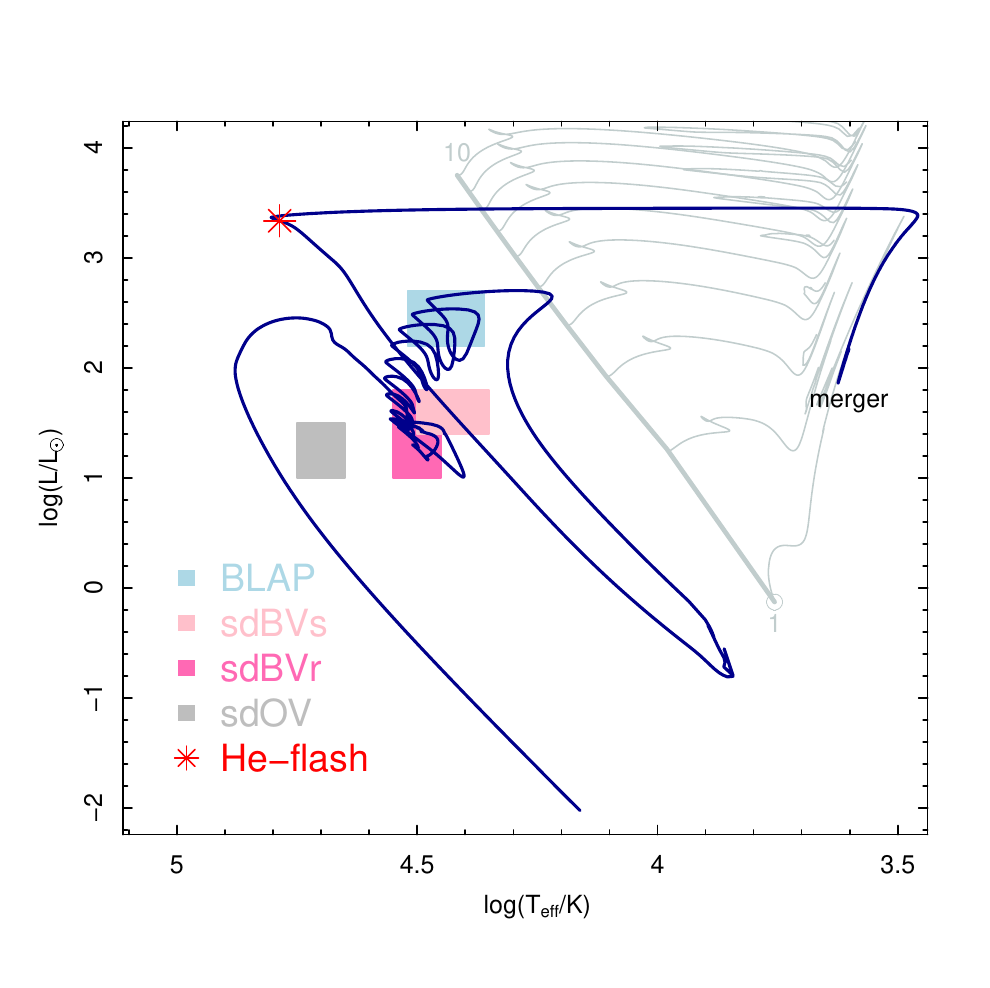}
\caption{Location of BLAPs and evolutionary tracks of HeWD+MS merger models in the Hertzsprung-Russell diagram.
The main sequence evolutionary tracks for 1 to 10 $\Msolar$ shown as grey lines for metallicity $Z = 0.02$ (mass fraction).
The black line indicates the evolutionary track of a $0.250$+$0.650\,\Msolar$ HeWD+MS merger remnant.
The star indicates the first helium shell flash (He-flash).
Square regions show locations of BLAPs and different types of pulsating hot subdwarfs, including short-period p-mode sdB stars (sdBVr: purple), long-period g-mode sdBs (sdBVs: pink), and the hotter sdO variables (sdOV: grey).  }
\label{1}
\end{figure}

Pulsating stars have been identified in diverse groups across the Hertzsprung--Russell (H--R) diagram \citep{Kurtz2022,Jeffery2016}.
Using periodic light variations,
asteroseismology is used to measure basic parameters and to infer the internal structure of pulsating stars \citep{Huber2011,Bedding2011,Chaplin2013}.
Due to the large scale-surveys,
more and more pulsating stars have been discovered.
Many new classes have been identified in recent years, including the blue large-amplitude pulsators (BLAPs) discovered by \cite{Pietrukowicz2017}.

BLAPs have several interesting properties:\\
(1) BLAPs have periods in the range 2 -- 60 min. As their name implies,
their light variations show a larger amplitude than other early-type pulsators with similar periods, i.e., from 0.05 to 0.4 mag at optical wavelengths.
In general, it is difficult for short-period stars to drive a high amplitude; e.g., the
pulsating Ap stars have periods in the range 5 -- 24 min and only have relatively lower amplitudes
from 0.001 to 0.02 mag. \\
(2) BLAPs have higher effective temperatures (25,000-34,000K) and
gravities ($\log (\rm g\, /cm\, s^{-2})$=4.2-5.7) than classical Cepheids and RR Lyrae-type stars but have similar saw-tooth light curves.\\
(3) The periods of BLAPs are not constant; a key observation is the detection of period changes
with both positive and negative signs observable on timescales of years \citep{Pietrukowicz2017}.
These imply that BLAPS are observed in both rapid expansion and contraction phases, which evolution models must be capable of explaining.\\
(4) Most BLAPs have an enrichment of surface helium  ($\log{n(\rm He)/n(\rm H)}$ in the range -2.4 to -0.4), which is similar to intermediate helium-rich hot subdwarfs.\\
(5) The observed BLAPs have been divided into two groups:
(a) low-gravity BLAPs (classical BLAPs) with pulsation periods in the range  20--60 mins, $\log (\rm g\, /cm\, s^{-2})$=4.2-4.7, and amplitudes 0.2--0.4 mag;
(b) high-gravity BLAPs, with periods in the range 2--8 mins, $\log (\rm g\, /cm\, s^{-2})$=5.3-5.7, and amplitudes 0.05-0.2 mag.
The high-gravity BLAPs reported by \cite{Kupfer2019} have spectral properties and pulsation periods similar to p-mode hot subdwarfs \citep{Ostensen2010}.
Unlike most BLAPs, TMTS-BLAP-1 is located in the period gap between low-gravity and high-gravity BLAPs\citep{Lin2022}.
BLAP OW-BLAP-1 is also found in the "gap"  \citep{Ramsay2022}.
Thus, the gap between low-gravity and high-gravity BLAPs may not be real.\\
(6) They are rare, and only HD 133729 has been clearly identified as binary \citep{Pigulski2022}.
Whether any others BLAPs are binaries is not known.

Thus, a successful model of BLAPs must explain the formation channel, the driving mechanism of pulsation, the
enrichment of helium abundance, the relation between low-gravity and high-gravity BLAPs, period changes and space density.
In the standard stellar evolution theory, obtaining a single model to explain all of these features is difficult.

Based on stellar structure, two principal scenarios to explain BLAPs were proposed by \cite{Pietrukowicz2017};
several authors have tried to reproduce the pulsational properties using models consistent with one or both scenarios.
(1) The shell-hydrogen-burning model: low-mass pre-white dwarfs ($\sim 0.3\,\Msolar$) with hydrogen-burning shells
pass through the BLAP instability zone, where the opacity of iron-group elements drives their pulsations  \citep{Wu2018,Byrne2018,Romero2018,Byrne2020,Byrne2021}.
(2) The helium-core burning model: BLAPs have similar effective temperatures but surface gravities lower than associated with the extended horizontal-branch.
Hence, BLAPs could be evolving toward or away from the extended horizontal-branch, either as pre- or post- hot subdwarfs
\citep{Wu2018,Kupfer2019,Meng2020,Xiong2022,Lin2022}.
Both models can represent some features of BLAPs, but not all.

On the Hertzsprung-Russell diagram, BLAPs are closely associated with the area occupied by hot subdwarfs.
Hot subdwarf stars can be roughly divided into subdwarf B (sdB), subdwarf O (sdO), and subdwarf OB (sdOB) by spectrum \citep{Heber2009,Heber2016}.
Most hot subdwarfs have a nearly pure hydrogen surface.
Some 10\% of hot subdwarfs have a surface helium abundance $> 90\%$ by number \citep{drilling13,Luo2016,Lei2019,Luo2021,Lei2022}.
The formation channel of most hot subdwarfs is well explained by the interaction of binaries \citep{Han2002,Han2003, Zhang12}.
Between the H-rich and He-rich type of stars, a small number of hot subdwarfs
have a surface helium number fraction of $10-90\%$ and are referred
to as intermediate helium-rich (iHe-rich) hot subdwarfs\citep{Ahmad03,Naslim2010,Jeffery2021}.
It is not known whether these represent an intermediate state of either H-rich or He-rich subdwarfs, some other evolution channel, or a combination of several channels.
For example, it is possible that they formed from a helium white dwarf (HeWD) merge with a main-sequence (MS) companion \citep{Zhang2017}.
BLAPS show a similar surface abundance to the intermediate helium-rich hot subdwarfs.
It is therefore helpful to investigate the properties and formation channels of iHe-rich subdwarfs as a possible channel for the formation of BLAPs.

One such channel is the white dwarf main sequence merger.
Many short-period detached binary systems consist of a helium white dwarf (HeWD) with a main-sequence (MS) companion \citep{Zorotovic2011}. For example,
\mbox{SDSS\,J121010.1+334722.9} is a cool $0.4\,\Msolar$ HeWD with a $0.16\,\Msolar$ M dwarf companion in a 3 h eclipsing binary.
Owing to a combination of gravitational-wave radiation, tidal interaction and magnetic braking, the orbital period and separation of such a binary can decrease over time and, as a consequence, the MS star may fill its Roche lobe.
If the MS star has a low mass, $\MMS \le 0.7\,\Msolar$, the mass transfer is expected to be dynamically unstable and lead to a merger \citep{Hurley2002,Shen2009}.
The immediate products of these HeWD+MS mergers are expected to be red giant branch-like (RGB-like) stars \citep{Hurley2002}.
Some such remnants are expected to ignite helium with a low envelope mass and thus become hot subdwarfs \citep{Clausen2011}.
\cite{Zhang2017}  found that some of the mergers result in the formation of hot subdwarfs with intermediate helium-rich surfaces.

In the \citet{Zhang2017} HeWD+MS merger model, the star takes a few tens of Myrs following the merger  before reaching the He-burning main sequence (or zero-age extended horizontal branch).
The evolutionary tracks pass through the region of the HR diagram occupied by BLAPs, and the surface abundance of helium, representing a mixture of hydrogen and helium, approximately matches that of BLAPs for which measurements exist.
This paper therefore examines the products of HeWD+MS mergers in more detail in order to assess their candidacy as BLAPs.
 In \S\,\ref{s_merger}, we introduce the post-merger
evolution models.
The comparison of theory with the observation of BLAPs
is shown in section \S\,\ref{s_results}. The conclusions
are in \S\,\ref{s_conclusion}.

\section{The mergers}

\begin{figure}
\plotone{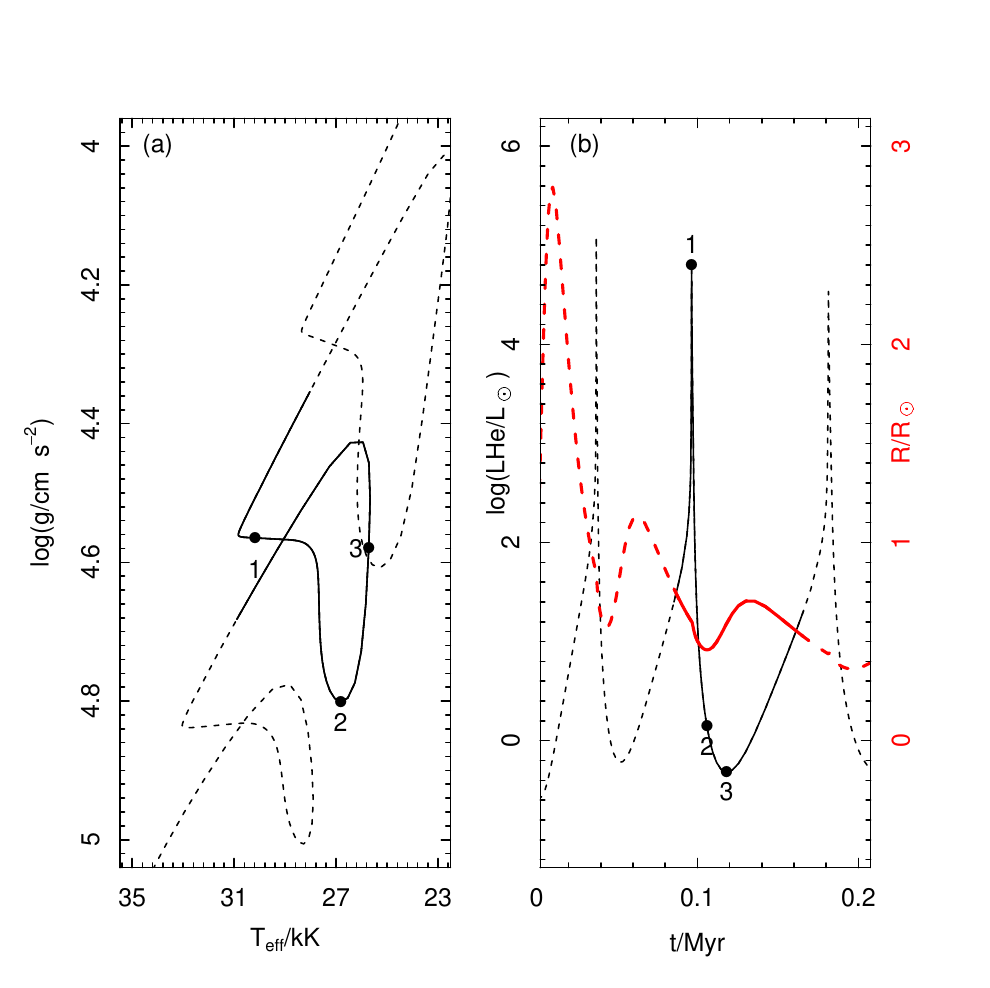}
\caption{A part of the post-merger evolution of a $0.250+0.650\,\Msolar$ HeWD+MS system during three helium shell flashes.
Left panel: The dashed line shows evolution in the $\rm log\,T_{\rm{eff}}$--$\log g$ diagram.
Solid line indicates a selected loop of evolution. Numbers indicate stages identified in the text.
Right panel: The evolution of radius (red) and helium-shell luminosity (black) for the same model.}
\label{2}
\end{figure}

\begin{figure}
\plotone{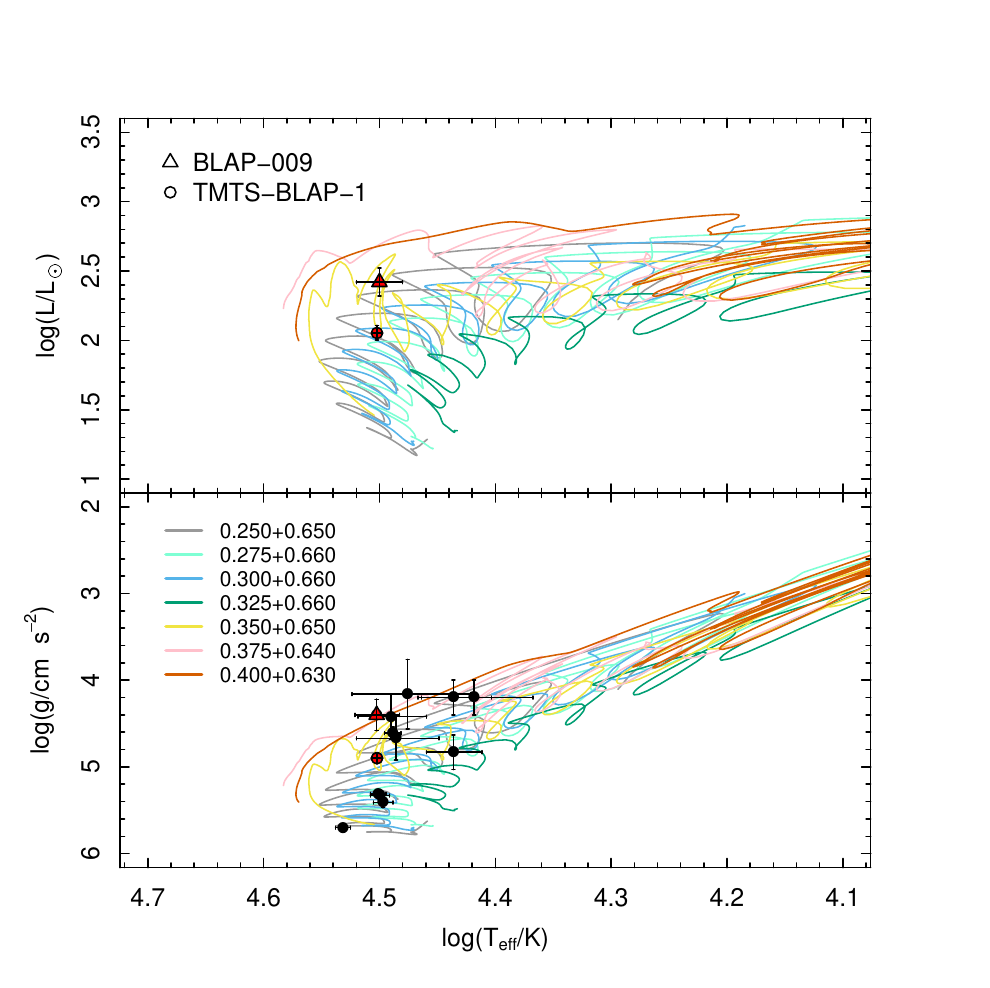}
\caption{Evolutionary tracks of seven HeWD+MS merger models that pass through parameter space where BLAPs are found. The tracks are distinguished by colour and identified by mass in the key. Top panel:
evolution in the $\rm log\,T_{\rm{eff}}$--$\log L$ (H-R) diagram. Bottom panel: evolution in the $\rm log\,T_{\rm{eff}}$--$\log g$ diagram.
Filled dots with errors represent observed BLAPs 1, 9, 11, 14 and 15 -- 23 from Table 1.
The triangle and circle show the stars with previously determined luminosity, OGLE-BLAP-009 \citep{Meng2020} and TMTS-BLAP-1 \citep{Lin2022}, respectively. }
\label{3}
\end{figure}

\begin{figure}
\plotone{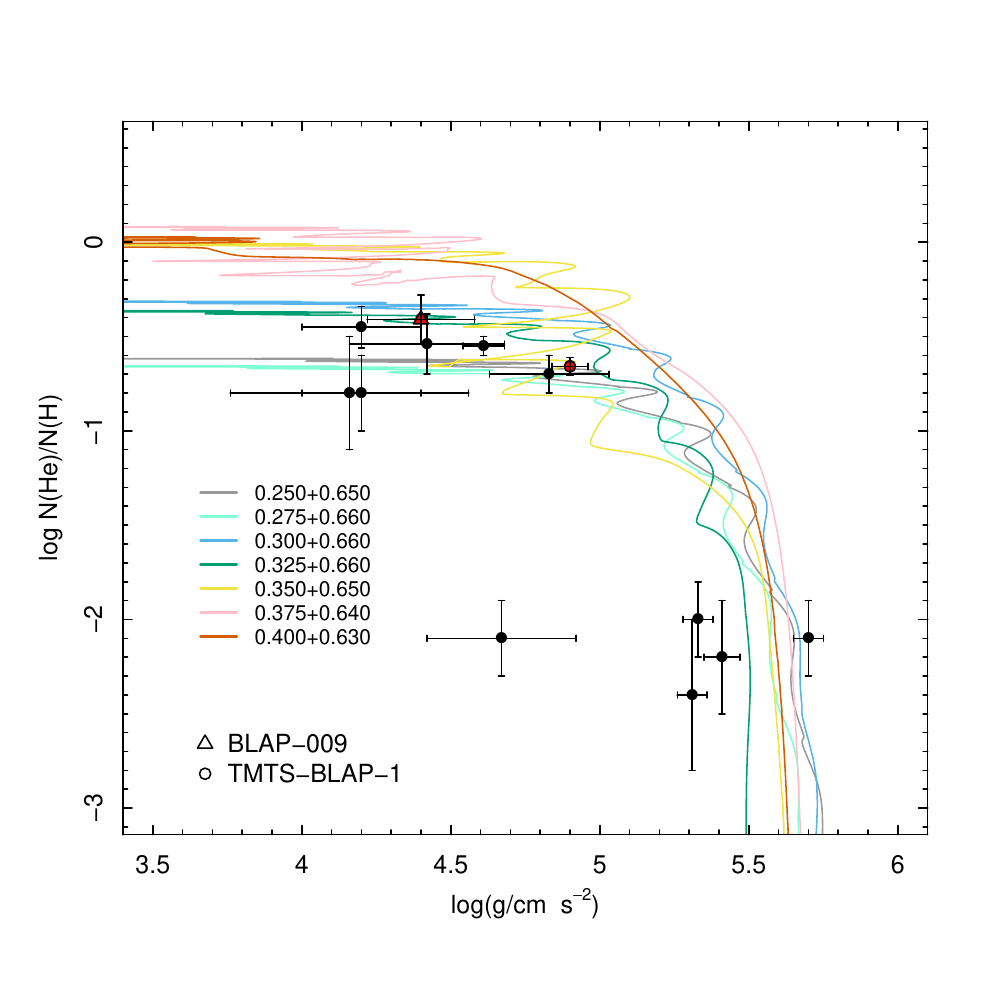}
\caption{Evolution of surface helium abundance $\log\,\rm N(He)/N(H)$ during helium shell flashes in seven possible BLAP models as a function of  $\log g$. Tracks are identified by colour and key as in Fig.\ref{3}.
Filled dots with errors represent observed BLAPs 1, 9, 11, 14 and 15 -- 23 in Table 1.  }
\label{4}
\end{figure}

\label{s_merger}
 \citet{Zhang2017} presented 29 HeWD+MS post-merger models. Some of the evolutionary tracks of post-merger pass through the region of the HR diagram occupied by BLAPs.
\citet{Zhang2017} have not analyzed the models for pulsation stability.
An opacity bump  due to iron-group element abundances  enhanced by radiation levitation is required to drive BLAP pulsations (e.g.,\citet{Byrne2020}).
It is therefore worth investigating the pulsation properties of the \citet{Zhang2017} post-merger models with full chemical diffusion.
We used the stellar evolution code {\sc mesa} version 23.05.11
\citep[Modules for Experiments in Stellar Astrophysics;][]{Paxton11, Paxton13, Paxton15, Paxton18, Paxton19} to calculate similar models to \citet{Zhang2017}.
We have analyzed the pulsation stability of each models using GYRE \citep{Townsend2013}.

\citet{Zhang2017} describe how, after the merger, an RGB-like star forms in which the structure is a very degenerate He core surrounded by an extended hydrogen envelope.
Following a similar evolution to normal RGB stars, hydrogen burns in a shell, and helium ash adds to the helium core.
Meanwhile mass is lost from the surface through a stellar wind.
Fresh helium continues to be produced during RGB-like evolution to compress and heat the He core.
Once the helium core is massive enough, the helium ignites in a shell,  followed by a series of helium flashes propagating inward toward the star center.
The first helium shell-flash is the strongest.
Its position near the surface of the degenerate He core drives a strong convection zone upwards that reaches the surface, enriching the surface in helium forced upward from beneath to create an iHe-sdB star.
The flash forces the star to expand, and then shrink as the shell luminosity drops again, so that the evolution track performs loops in the HR diagram \citep{Sweigart1997,Brown2001,Saio2000}.
After a few Myrs, the He-burning flame reaches the center and the star commences true core He burning, i.e., on the zero-age extended horizontal branch.
While helium burns in the center, heavier elements near the surface diffuse downwards and finally produce an almost pure hydrogen atmosphere.
Thus, the remnants evolve to become hydrogen-rich single hot subdwarfs (H-sdB).
The entire evolutionary track might be represented as:
$\rm RGB\rightarrow  \rm iHe-sdB \,(BLAP)\rightarrow \rm H-sdB\rightarrow WD$

As identified by \citet{Zhang2017}, the evolution of HeWD+MS mergers can be divided into two paths corresponding to an {\it early} hot flasher or a {\it late} hot flasher \citep{Lanz2004,Heber2009,Heber2016}.
(1) late hot flasher: the hydrogen envelope is of very low mass, so flash-driven convection can yield a maximum helium surface abundance $Y=0.954$ (mass fraction);
(2) early hot flasher: the hydrogen shell re-ignites after the helium shell flash, the star expands, and initiates deep opacity-driven surface convection. The combination of flash-driven convection followed by opacity-driven convection some  helium and other newly produced elements are dredged to the surface, yielding a maximum surface abundance $Y=0.636$.

Analyzing each model in detail, we selected seven early hot flasher models as possible BLAPs, by having tracks which pass through a similar volume of gravity-temperature space and $\log{n(\rm He)/n(\rm H)} \le 0$. These models are, $0.250$+$0.650\,\Msolar$,
$0.275$+$0.660\,\Msolar$, $0.300$+$0.660\,\Msolar$, $0.325$+$0.660\,\Msolar$,
$0.350$+$0.650\,\Msolar$, $0.375$+$0.650\,\Msolar$ and $0.400$+$0.640\,\Msolar$, where the first and second quantity in each pair refers to the progenitor HeWD and MS mass, respectively.
The masses for each model as they cross the BLAP region are $0.484$, $0.492$, $0.499$, $0.511$, $0.517$, $0.527$ and $0.543\Msolar$.
Thus, the masses of BLAPs are in the range $0.484-0.543\Msolar$.

Fig.~\ref{1} shows a detailed post-merger evolutionary track of an early hot flasher model, i.e., $0.250 + 0.650\,\Msolar$ remnant.
The flash-driven loops span the region where BLAPs are observed and then reach the zone of sdB stars.
Fig.~\ref{2} shows an expanded section of the same evolutionary track. Part of flash loop is
marked by points 1 to 3, which represent key phases during the loop including
(1) peak of the helium flash, (2) minimum radius, and (3) He-shell luminosity minimum. The duration of the helium shell flash is short. It is accompanied initially by a halt in the envelope contraction (1) and a drop in total luminosity (1)-(2), followed by envelope expansion as heat generated in the flash is transmitted outward on a thermal timescale (2)-(3). Once the envelope reaches radiative equilibrium after the flash, contraction will resume at a total luminosity slightly less than before the flash.

\begin{table*}
\caption{List of BLAPs.}
 \label{table 1}
\centering
\begin{tabular}{lccccccc}     
\hline\hline
 No.& name & $p({\rm min})$& $\dot{p}/p(10^{-7}\rm yr^{-1})$ &$T_{\rm eff}$  &$\log{g}$&$\log{n(\rm He)/n(\rm H)}$& Ref. \\
 \hline
  1& OGLE-BLAP-001 &28.26 &2.90$\pm$3.70 &30800$\pm$500&4.61$\pm$0.07 &$-0.55\pm$0.05& (1) \\
  2& OGLE-BLAP-002  &23.29 &$-19.23\pm$8.05  & -&- &-& (1) \\
  3& OGLE-BLAP-003  &28.46 &0.82$\pm$0.32 &- &- & -& (1) \\
  4& OGLE-BLAP-004  &22.36 &$-5.03\pm$1.57 &- & -& -& (1) \\
  5& OGLE-BLAP-005  &27.25 &0.63$\pm$0.26 &- &- &-& (1) \\
  6& OGLE-BLAP-006  &38.02 &$-2.85\pm$0.31 &- &- &- & (1) \\
  7& OGLE-BLAP-007  &35.18 &$-2.40\pm$0.51 &- &- &- & (1) \\
  8& OGLE-BLAP-008  &34.48 &2.11$\pm$0.27 &- &- &-& (1) \\
  9& OGLE-BLAP-009  &31.94 &1.63$\pm$0.08 &31800$\pm$1400&4.40$\pm$0.18 &$-0.41\pm$0.13& (1) \\
  10& OGLE-BLAP-010  &32.13 &0.44$\pm$0.21 &- &- &-& (1) \\
  11& OGLE-BLAP-011  &34.87 &6.77$\pm$8.87 &26200$\pm$2900&4.20$\pm$0.20 &$-0.45\pm$0.11& (1) \\
  12& OGLE-BLAP-012  &30.90 &0.03$\pm$0.15 &- &- &-& (1) \\
  13& OGLE-BLAP-013  &39.33 &7.65$\pm$0.67 &- &- &-& (1) \\
  14& OGLE-BLAP-014  &33.62 &4.82$\pm$0.39 &30900$\pm$2100&4.42$\pm$0.26 &$-0.54\pm$0.16& (1) \\
  15& high-gravity-BLAP-1& 3.34& -& 34000$\pm$500& 5.70$\pm$0.05&-2.1$\pm$0.2& (2) \\
  16& high-gravity-BLAP-2& 6.05& -& 31400$\pm$600& 5.41$\pm$0.06&-2.2$\pm$0.3& (2) \\
  17& high-gravity-BLAP-3& 7.31& -& 31600$\pm$600& 5.33$\pm$0.05&-2.0$\pm$0.2& (2) \\
  18& high-gravity-BLAP-4& 7.92& -& 31700$\pm$500& 5.31$\pm$0.05&-2.4$\pm$0.4& (2) \\
  19& OW-BLAP-1& 10.8& -& 30600$\pm$2500& 4.67$\pm$0.25& $-2.1\pm$0.2& (3) \\
  20& OW-BLAP-2& 23.0& -& 27300$\pm$1500& 4.83$\pm$0.20& $-0.7\pm$0.1& (3) \\
  21& OW-BLAP-3& 28.9& -& 29900$\pm$3500& 4.16$\pm$0.40& $-0.8\pm$0.3& (3) \\
  22& OW-BLAP-4& 32.0& -& 27300$\pm$2000& 4.20$\pm$0.20& $-0.8\pm$0.2& (3) \\
  23& TMTS-BLAP-1 	&18.9 & 22.3$\pm$0.9&31780$\pm$350& 4.90$\pm$0.06&$-0.66\pm$0.05& (4) \\
  24& HD 133729  &32.27 &$-11.5$ & 29000 & 4.5 & -& (5)\\
 \hline\hline
\end{tabular}
\begin{minipage}{1\textwidth}
(1) \cite{Pietrukowicz2017}, (2) \cite{Kupfer2019}, (3) \cite{Ramsay2022}, (4) \cite{Lin2022},  (5) \cite{Pigulski2022}.
\end{minipage}
\end{table*}

\section{Comparison with observation}
\label{s_results}

Having found that model HeWD+MS merger remnants can become BLAPs, we compare their
properties to observed examples of such stars in more detail.
Table 1 shows a sample of 24 confirmed BLAPs \citep{Pietrukowicz2017,Kupfer2019,Lin2022,Pigulski2022,Ramsay2022}.
Since our model is for merged stars, {\bf which, unless they were originally in a triple or higher multiplicity system, should now be single}, we do not including the binary HD 133729 for comparison \citep{Pigulski2022}.
Fig.~\ref{3} shows these observed BLAPs
and the tracks for the seven possible BLAP models in both the $\rm log T_{\rm{eff}}$--$\log L$ and $\rm log T_{\rm{eff}}$--$\log g$ planes.
Fig.~\ref{4} compares models and observations in the $\log g$-
surface helium abundance plane.
This figure shows that the
model HeWD+MS merger remnants can explain the surface helium abundance of BLAPs, with a strong preference for the lowest mass models:
$0.250$+$0.650\,\Msolar$,
$0.275$+$0.660\,\Msolar$,
$0.300$+$0.660\,\Msolar$,
$0.325$+$0.660\,\Msolar$ and
$0.350$+$0.650\,\Msolar$.
Fig.~\ref{3} and Fig.~\ref{4} show that the spaces occupied by low-gravity and high-gravity BLAPs
represent different stages of similar tracks, which indicates a possible evolutionary relation between both types of BLAP.
The TMTS-BLAP-1 and  OW-BLAP-1 could also also associated with the evolution of merger models.
At least, our theoretical models have no ``gap'' between low-gravity and high-gravity BLAPs.

\begin{figure}
\plotone{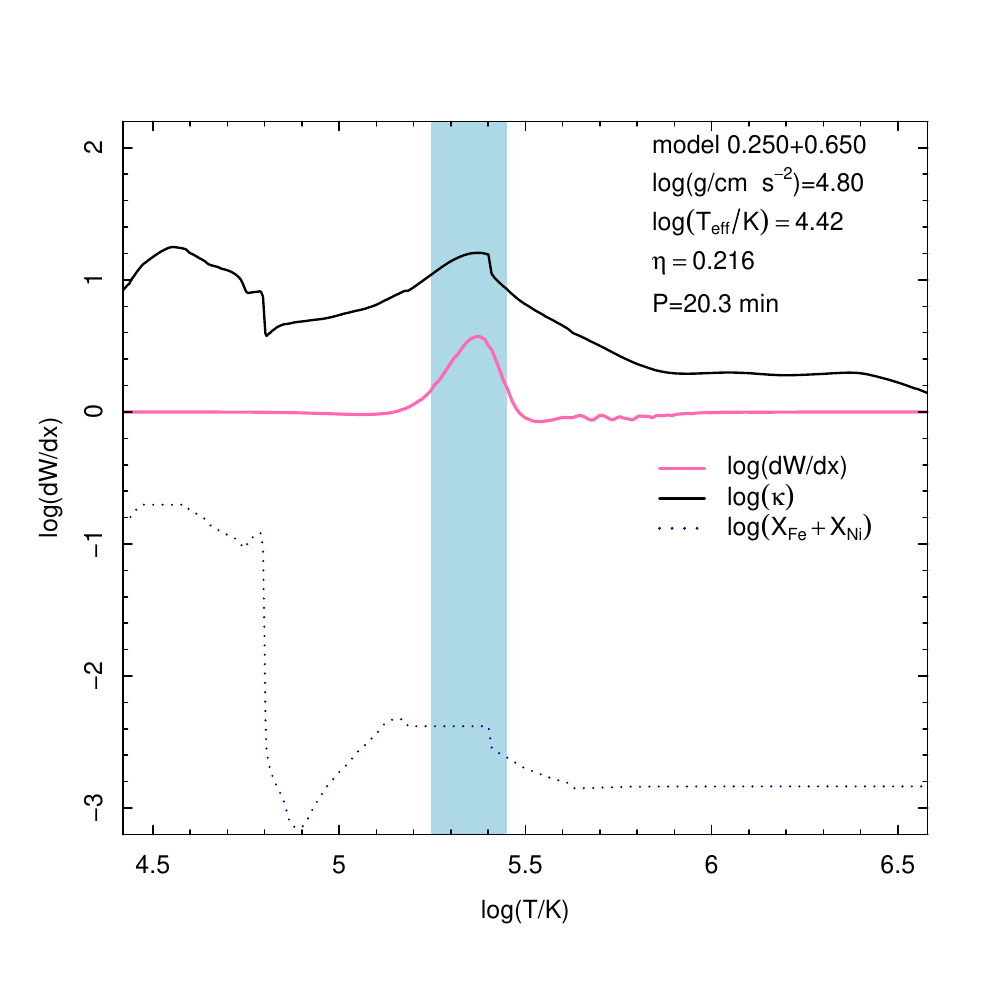}
\caption{Profile of opacity, the mass fraction of Fe and Ni and work function
dW/dx of stellar interior temperature. The logarithm of the opacity is shown by solid
black line, the logarithm of the mass fraction of Fe and Ni is indicated by the dotted line, while the
value of dW/dx is indicated by the solid pink line. The blue zone indicate the approximate temperatures of the partial ionisation
opacity peaks of iron group elements.
}
\label{5}
\end{figure}

\begin{figure}
\plotone{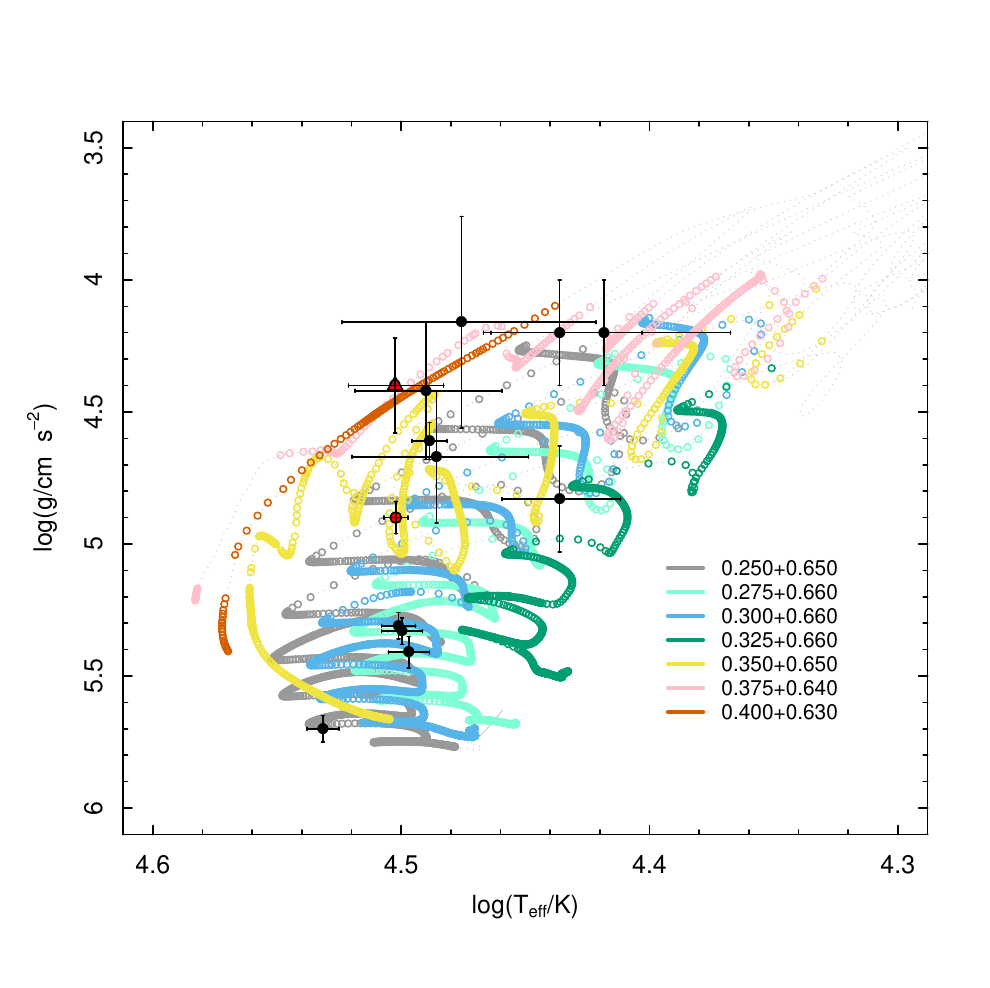}
\caption{The pulsation stability of the seven HeWD+MS merger models on the $\rm log T_{\rm{eff}}$--$\log g$ plane.
The tracks are indicated by grey dotted lines, while the unstable modes are shown by coloured circles and and identifed in the key.
The BLAPs shown in Fig.~\ref{3} are also indicated. }
\label{6}
\end{figure}

\begin{figure}
\plotone{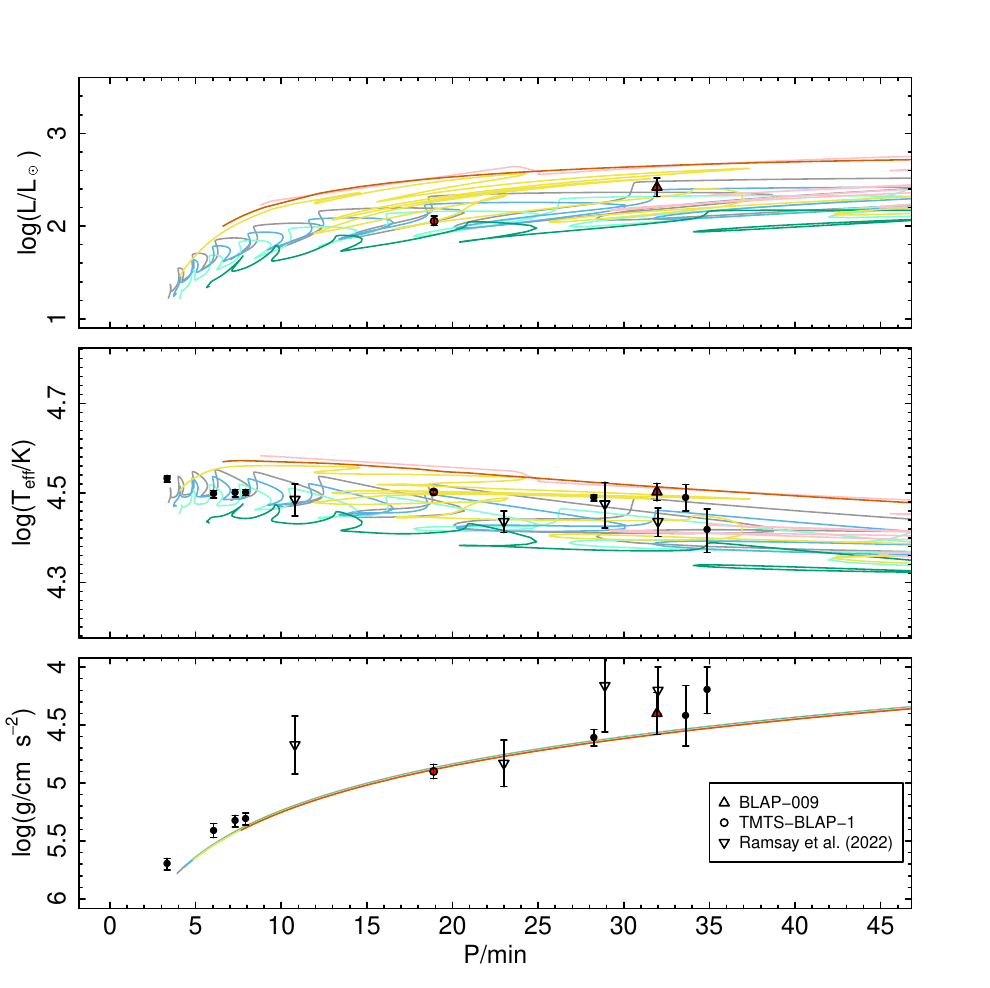}
\caption{Evolution of 7 possible BLAP models in the period-luminosiuty (top), period-temperature (middle) and period-gravity (bottom) planes. Tracks are identified by colour as in Fig.\ref{3}. Filled dots with errors represent observed BLAPs from Table 1.
}
\label{7}
\end{figure}

\begin{figure}
\plotone{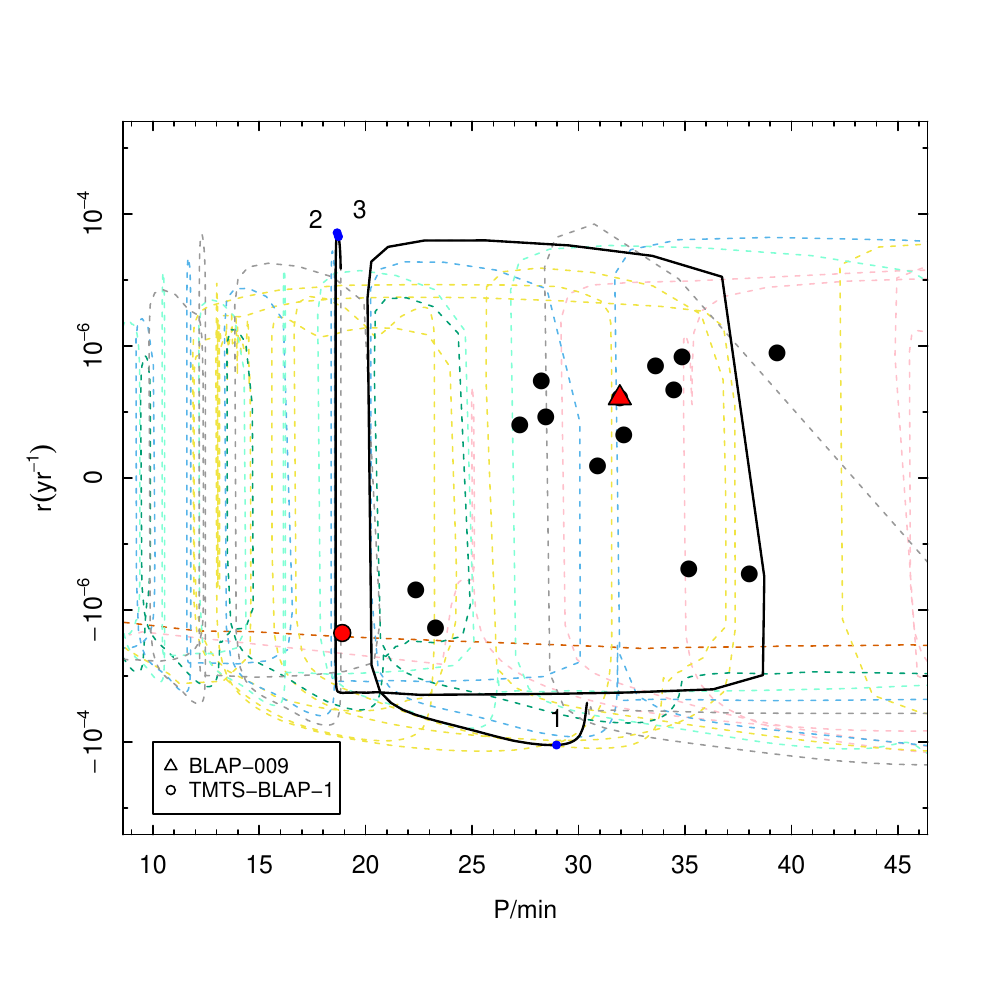}
\caption{The rate of period change $r\equiv\dot{P}/P$ versus pulsation period $P$ for the seven possible BLAP models.
Tracks are identified by colour as in Fig.\ref{3}.
The solid line shows the same loop through a helium shell flash for the $0.250+0.650$\,\Msolar\ post-merger model shown in Fig.~\ref{2}.
Filled dots represent BLAPs  1--14 and 23 in Table 1.}
\label{8}
\end{figure}

We use the oscillation code GYRE to analyze the stability of the models.
The input model for GYRE is obtained from the MESA calculation and then uses a non-adiabatic analysis to investigate the stability of each mode.
We only calculate the stability ($\eta$) and period (P) for radial modes ($l = 0$) as the BLAPs are known
to be large amplitude pulsators. Modes with $\eta>0$ are unstable.
Fig.~\ref{5} shows the logarithm of opacity (log($\kappa$)) and
the work function ($\rm dW/dx$) as a function of interior temperature for a model around point 2 of Fig.~\ref{2}. The mass fraction of iron and nickel are also included.
The location of the peak in $\rm dW/dx$ coincides with the area of maximum opacity, which is related to the ionization of iron and nickel
around $\rm log(T/K)=5.35$. Thus, this unstable mode is driven
by the $\kappa$-mechanism related to the opacity bump due to the accumulation of iron-group elements in the envelope.
We also check the models without radiative levitation, which modes are always stable.
These results agree with previous studies
that the modes are only excited when radiative levitation is included \citep{Byrne2018,Romero2018,Byrne2020}.
Fig.~\ref{6} shows the pulsation stability of the seven models. The locations of unstable models are shown by colored circles.
Sections of the tracks shown in Fig.~\ref{3} which are stable against pulsation are shown as dotted lines.
The instability region satisfactorily coincides with the region where BLAPS are observed and  indicates why BLAPS might not be observed outside this region.

BLAPs are considered to be rapidly evolving stars.
Both the pulsation period and the rates of period change
are important features to compare with observation. We calculate the period $P$ and the corresponding rates of period change $r$ for the fundamental radial mode from our models.
Comparing $P$, $M$ and $R$, the classical period mean-density relation \citep{Eddington1918} for radial-mode pulsations,
\begin{equation}
P \approx Q_{\rm F} / \sqrt{\frac{M}{M_\odot}  \left(\frac{R_\odot}{R}\right)^3} \,{\rm min},
\end{equation}
holds with the pulsation constant for the fundamental radial mode $Q_{\rm F} \approx 47$\,min.

Fig.~\ref{7} shows the observed periods compared with models in the $P - \log T_{\rm{eff}}$ and $P - \log g$ planes.
In the $P - \log T_{\rm{eff}}$ plane,  the locations of all observed BLAPs are in the region of our proposed 7 possible merger models.
Meanwhile, the same models can explain all of the BLAPs in the $P - \log g$ plane; six out of thirteen lie within 1$\sigma$ (68\% confidence interval), twelve lie within 2$\sigma$ (95\%) and all lie within 3$\sigma$ (99\%) of the model predictions.

Using the same models, we calculate the expected period change $r_i$ from
\begin{equation}
r_i \equiv \frac{\dot{P}}{P} = \frac{\Delta P}{\Delta t} \frac{1}{P} = \frac{P_{i+1}-P_{i}}{t_{i+1}-t_{i}} \frac{1}{P_{i+1}},
\end{equation}
where $i$ refers to each model along a track.
Fig.~\ref{8} shows the rates of period change in the $P - r$ plane.
It is a feature of post-merger models that, like BLAPs, they show both positive and negative values of $r$ as a consequence of the $r-T_{\rm eff}$ loops associated with helium shell flashes.
The 15 BLAPs with measured $r$ all lie within the range covered by the theoretical  values for the 7 possible BLAP models: $\sim -10^{-5} \,\rm yr^{-1}$ to $\sim 10^{-5} \,\rm yr^{-1}$.
Fig.~\ref{8} highlights the loop covering one helium-shell flash for the $0.250+0.650$\,\Msolar\ post-merger model already shown on the $\log T_{\rm{eff}} - \log g$ plane in Fig.~\ref{2}.
This demonstrates how period decreases (negative $\dot{P}$) as the star contracts, but briefly increases (positive $\dot{P}$) during the shell-flash itself.

From Fig.~\ref{3} to Fig.~\ref{8}, most properties of BLAPs can be reasonably well
reproduced by our merger model. Furthermore,  the TMTS-BLAP-1 was identified as the ``Hertzsprung Gap'' of Hot Subdwarfs \citep{Lin2022}.
Here, we suggest that TMTS-BLAP-1 also formed from a merger and is currently in the pre-sdB stage, as shown in our models.

\section{Conclusion}
\label{s_conclusion}
From an analysis of the HeWD+MS post-merger models, we have found that some of the early-flasher models provide a possible channel for the origin of BLAPs.
Such mergers had previously been identified as a channel to form iHe-rich hot subdwarfs, and that is confirmed here.
Some of these models transit the  parameter space occupied by BLAPs during their helium-shell flash phase prior to full helium core burning.
At this point a star represented by these models will become an iHe-rich sdB star and eventually, the heavier elements will sink to leave an H-rich surface as the star approaches the He-burning main sequence as a single H-rich sdB star.

Analysis of the distribution of post-mergers evolution tracks
shows that predictions for HeWD+MS mergers are consistent with observations of almost all recent BLAPs in terms of
surface effective temperature ($T_{\rm eff}$), surface gravity ($\rm log\, g$),
surface luminosity ($\rm log\, L$), surface helium abundance, period ($P$), and rates of period change ($r$).
Significantly, because of the cyclic expansion and contraction of helium-shell-flashing stars, the models predict both negative and positive values for the period change, as observed.

BLAPS that evolve from mergers are likely to be single stars. It may be that BLAPs, especially BLAPs in binaries, can also be formed in other channels
or formed in triple systems and leave behind a merged star in a binary\citep{Preece2022}.
Recently, the BLAP HD133729 was found to be binary. We hope that further observations will identify whether any other BLAPs have companions.

\begin{acknowledgments}
We thank the referee for the helpful suggestions
and comments that improved the manuscript.
This work is supported by the grants 12073006, 12090040,12090042,12288102,12133011 and 11833006 from the National Natural Science Foundation of China,
the Joint Research Fund in Astronomy (U2031203) under cooperative agreement between
the National Natural Science Foundation of China (NSFC) and Chinese Academy of Sciences (CAS)
and the CAS "Light of West China" Program.
We also acknowledge the science research grant from the China Manned Space Project with No.CMS-CSST-2021-A10.
Armagh Observatory and Planetarium is supported by a grant from the Northern Ireland Department for Communities.
X.Z. thanks Jie Lin and Tao Wu for helpful conversations.
\end{acknowledgments}

\bibliographystyle{aasjournal} 
\bibliography{mybib} 

\begin{thebibliography}{}
\expandafter\ifx\csname natexlab\endcsname\relax\def\natexlab#1{#1}\fi
\providecommand{\url}[1]{\href{#1}{#1}}
\providecommand{\dodoi}[1]{doi:~\href{http://doi.org/#1}{\nolinkurl{#1}}}
\providecommand{\doeprint}[1]{\href{http://ascl.net/#1}{\nolinkurl{http://ascl.net/#1}}}
\providecommand{\doarXiv}[1]{\href{https://arxiv.org/abs/#1}{\nolinkurl{https://arxiv.org/abs/#1}}}

\bibitem[{{Ahmad} \& {Jeffery}(2003)}]{Ahmad03}
{Ahmad}, A., \& {Jeffery}, C.~S. 2003, A\&A, 402, 335,
  \dodoi{10.1051/0004-6361:20030233}

\bibitem[{{Bedding} {et~al.}(2011){Bedding}, {Mosser}, {Huber},
  {Montalb{\'a}n}, {Beck}, {Christensen-Dalsgaard}, {Elsworth}, {Garc{\'\i}a},
  {Miglio}, {Stello}, {White}, {De Ridder}, {Hekker}, {Aerts}, {Barban},
  {Belkacem}, {Broomhall}, {Brown}, {Buzasi}, {Carrier}, {Chaplin}, {di Mauro},
  {Dupret}, {Frandsen}, {Gilliland}, {Goupil}, {Jenkins}, {Kallinger},
  {Kawaler}, {Kjeldsen}, {Mathur}, {Noels}, {Silva Aguirre}, \&
  {Ventura}}]{Bedding2011}
{Bedding}, T.~R., {Mosser}, B., {Huber}, D., {et~al.} 2011, Nature, 471, 608,
  \dodoi{10.1038/nature09935}

\bibitem[{{Brown} {et~al.}(2001){Brown}, {Sweigart}, {Lanz}, {Landsman}, \&
  {Hubeny}}]{Brown2001}
{Brown}, T.~M., {Sweigart}, A.~V., {Lanz}, T., {Landsman}, W.~B., \& {Hubeny},
  I. 2001, ApJ, 562, 368, \dodoi{10.1086/323862}

\bibitem[{{Byrne} \& {Jeffery}(2018)}]{Byrne2018}
{Byrne}, C.~M., \& {Jeffery}, C.~S. 2018, MNRAS, 481, 3810,
  \dodoi{10.1093/mnras/sty2545}

\bibitem[{{Byrne} \& {Jeffery}(2020)}]{Byrne2020}
---. 2020, MNRAS, 492, 232, \dodoi{10.1093/mnras/stz3486}

\bibitem[{{Byrne} {et~al.}(2021){Byrne}, {Stanway}, \& {Eldridge}}]{Byrne2021}
{Byrne}, C.~M., {Stanway}, E.~R., \& {Eldridge}, J.~J. 2021, MNRAS, 507, 621,
  \dodoi{10.1093/mnras/stab2115}

\bibitem[{{Chaplin} \& {Miglio}(2013)}]{Chaplin2013}
{Chaplin}, W.~J., \& {Miglio}, A. 2013, ARA\&A, 51, 353,
  \dodoi{10.1146/annurev-astro-082812-140938}

\bibitem[{{Clausen} \& {Wade}(2011)}]{Clausen2011}
{Clausen}, D., \& {Wade}, R.~A. 2011, ApJL, 733, L42,
  \dodoi{10.1088/2041-8205/733/2/L42}

\bibitem[{{Drilling} {et~al.}(2013){Drilling}, {Jeffery}, {Heber}, {Moehler},
  \& {Napiwotzki}}]{drilling13}
{Drilling}, J.~S., {Jeffery}, C.~S., {Heber}, U., {Moehler}, S., \&
  {Napiwotzki}, R. 2013, \aap, 551, A31, \dodoi{10.1051/0004-6361/201219433}

\bibitem[{{Eddington}(1918)}]{Eddington1918}
{Eddington}, A.~S. 1918, MNRAS, 79, 2

\bibitem[{{Han} {et~al.}(2003){Han}, {Podsiadlowski}, {Maxted}, \&
  {Marsh}}]{Han2003}
{Han}, Z., {Podsiadlowski}, P., {Maxted}, P.~F.~L., \& {Marsh}, T.~R. 2003,
  MNRAS, 341, 669, \dodoi{10.1046/j.1365-8711.2003.06451.x}

\bibitem[{{Han} {et~al.}(2002){Han}, {Podsiadlowski}, {Maxted}, {Marsh}, \&
  {Ivanova}}]{Han2002}
{Han}, Z., {Podsiadlowski}, P., {Maxted}, P.~F.~L., {Marsh}, T.~R., \&
  {Ivanova}, N. 2002, MNRAS, 336, 449, \dodoi{10.1046/j.1365-8711.2002.05752.x}

\bibitem[{{Heber}(2009)}]{Heber2009}
{Heber}, U. 2009, ARA\&A, 47, 211, \dodoi{10.1146/annurev-astro-082708-101836}

\bibitem[{{Heber}(2016)}]{Heber2016}
---. 2016, PASP, 128, 082001, \dodoi{10.1088/1538-3873/128/966/082001}

\bibitem[{{Huber} {et~al.}(2011){Huber}, {Bedding}, {Stello}, {Hekker},
  {Mathur}, {Mosser}, {Verner}, {Bonanno}, {Buzasi}, {Campante}, {Elsworth},
  {Hale}, {Kallinger}, {Silva Aguirre}, {Chaplin}, {De Ridder}, {Garc{\'\i}a},
  {Appourchaux}, {Frandsen}, {Houdek}, {Molenda-{\.Z}akowicz}, {Monteiro},
  {Christensen-Dalsgaard}, {Gilliland}, {Kawaler}, {Kjeldsen}, {Broomhall},
  {Corsaro}, {Salabert}, {Sanderfer}, {Seader}, \& {Smith}}]{Huber2011}
{Huber}, D., {Bedding}, T.~R., {Stello}, D., {et~al.} 2011, ApJ, 743, 143,
  \dodoi{10.1088/0004-637X/743/2/143}

\bibitem[{{Hurley} {et~al.}(2002){Hurley}, {Tout}, \& {Pols}}]{Hurley2002}
{Hurley}, J.~R., {Tout}, C.~A., \& {Pols}, O.~R. 2002, MNRAS, 329, 897,
  \dodoi{10.1046/j.1365-8711.2002.05038.x}

\bibitem[{{Jeffery} {et~al.}(2021){Jeffery}, {Miszalski}, \&
  {Snowdon}}]{Jeffery2021}
{Jeffery}, C.~S., {Miszalski}, B., \& {Snowdon}, E. 2021, MNRAS, 501, 623,
  \dodoi{10.1093/mnras/staa3648}

\bibitem[{{Jeffery} \& {Saio}(2016)}]{Jeffery2016}
{Jeffery}, C.~S., \& {Saio}, H. 2016, MNRAS, 458, 1352,
  \dodoi{10.1093/mnras/stw388}

\bibitem[{{Kupfer} {et~al.}(2019){Kupfer}, {Bauer}, {Burdge}, {Bellm},
  {Bildsten}, {Fuller}, {Hermes}, {Kulkarni}, {Prince}, {van Roestel},
  {Dekany}, {Duev}, {Feeney}, {Giomi}, {Graham}, {Kaye}, {Laher}, {Masci},
  {Porter}, {Riddle}, {Shupe}, {Smith}, {Soumagnac}, {Szkody}, \&
  {Ward}}]{Kupfer2019}
{Kupfer}, T., {Bauer}, E.~B., {Burdge}, K.~B., {et~al.} 2019, ApJL, 878, L35,
  \dodoi{10.3847/2041-8213/ab263c}

\bibitem[{{Kurtz}(2022)}]{Kurtz2022}
{Kurtz}, D.~W. 2022, ARA\&A, 60, 31,
  \dodoi{10.1146/annurev-astro-052920-094232}

\bibitem[{{Lanz} {et~al.}(2004){Lanz}, {Brown}, {Sweigart}, {Hubeny}, \&
  {Landsman}}]{Lanz2004}
{Lanz}, T., {Brown}, T.~M., {Sweigart}, A.~V., {Hubeny}, I., \& {Landsman},
  W.~B. 2004, ApJ, 602, 342, \dodoi{10.1086/380904}

\bibitem[{{Lei} {et~al.}(2019){Lei}, {Zhao}, {N{\'e}meth}, \& {Zhao}}]{Lei2019}
{Lei}, Z., {Zhao}, J., {N{\'e}meth}, P., \& {Zhao}, G. 2019, ApJ, 881, 135,
  \dodoi{10.3847/1538-4357/ab2edc}

\bibitem[{{Lei} {et~al.}(2022){Lei}, {He}, {Nemeth}, {Vos}, {Zou}, {Hu},
  {Xiao}, {Yan}, \& {Zhao}}]{Lei2022}
{Lei}, Z., {He}, R., {Nemeth}, P., {et~al.} 2022, arXiv e-prints,
  arXiv:2211.12323.
\newblock \doarXiv{2211.12323}

\bibitem[{{Lin} {et~al.}(2022){Lin}, {Wu}, {Wang}, {N{\'e}meth}, {Xiong}, {Wu},
  {Filippenko}, {Cai}, {Brink}, {Yan}, {Zeng}, {Luo}, {Xiang}, {Zhang},
  {Zheng}, {Yang}, {Mo}, {Xi}, {Zhang}, {Iskandar}, {Esamdin}, {Jiang}, {Sai},
  {Wei}, {Chen}, {Guo}, {Chen}, {Li}, {Lin}, {Lin}, \& {Zhang}}]{Lin2022}
{Lin}, J., {Wu}, C., {Wang}, X., {et~al.} 2022, Nature Astronomy,
  \dodoi{10.1038/s41550-022-01783-z}

\bibitem[{{Luo} {et~al.}(2021){Luo}, {N{\'e}meth}, {Wang}, {Wang}, \&
  {Han}}]{Luo2021}
{Luo}, Y., {N{\'e}meth}, P., {Wang}, K., {Wang}, X., \& {Han}, Z. 2021, ApJS,
  256, 28, \dodoi{10.3847/1538-4365/ac11f6}

\bibitem[{{Luo} {et~al.}(2016){Luo}, {N{\'e}meth}, {Liu}, {Deng}, \&
  {Han}}]{Luo2016}
{Luo}, Y.-P., {N{\'e}meth}, P., {Liu}, C., {Deng}, L.-C., \& {Han}, Z.-W. 2016,
  ApJ, 818, 202, \dodoi{10.3847/0004-637X/818/2/202}

\bibitem[{{Meng} {et~al.}(2020){Meng}, {Han}, {Podsiadlowski}, \&
  {Li}}]{Meng2020}
{Meng}, X.-C., {Han}, Z.-W., {Podsiadlowski}, P., \& {Li}, J. 2020, ApJ, 903,
  100, \dodoi{10.3847/1538-4357/abbb8e}

\bibitem[{{Naslim} {et~al.}(2010){Naslim}, {Jeffery}, {Ahmad}, {Behara}, \&
  {{\c{S}}ah{\`\i}n}}]{Naslim2010}
{Naslim}, N., {Jeffery}, C.~S., {Ahmad}, A., {Behara}, N.~T., \&
  {{\c{S}}ah{\`\i}n}, T. 2010, MNRAS, 409, 582,
  \dodoi{10.1111/j.1365-2966.2010.17324.x}

\bibitem[{{{\O}stensen} {et~al.}(2010){{\O}stensen}, {Oreiro}, {Solheim},
  {Heber}, {Silvotti}, {Gonz{\'a}lez-P{\'e}rez}, {Ulla}, {P{\'e}rez
  Hern{\'a}ndez}, {Rodr{\'\i}guez-L{\'o}pez}, \& {Telting}}]{Ostensen2010}
{{\O}stensen}, R.~H., {Oreiro}, R., {Solheim}, J.~E., {et~al.} 2010, A\&A, 513,
  A6, \dodoi{10.1051/0004-6361/200913480}

\bibitem[{{Paxton} {et~al.}(2011){Paxton}, {Bildsten}, {Dotter}, {Herwig},
  {Lesaffre}, \& {Timmes}}]{Paxton11}
{Paxton}, B., {Bildsten}, L., {Dotter}, A., {et~al.} 2011, ApJS, 192, 3,
  \dodoi{10.1088/0067-0049/192/1/3}

\bibitem[{{Paxton} {et~al.}(2013){Paxton}, {Cantiello}, {Arras}, {Bildsten},
  {Brown}, {Dotter}, {Mankovich}, {Montgomery}, {Stello}, {Timmes}, \&
  {Townsend}}]{Paxton13}
{Paxton}, B., {Cantiello}, M., {Arras}, P., {et~al.} 2013, ApJS, 208, 4,
  \dodoi{10.1088/0067-0049/208/1/4}

\bibitem[{{Paxton} {et~al.}(2015){Paxton}, {Marchant}, {Schwab}, {Bauer},
  {Bildsten}, {Cantiello}, {Dessart}, {Farmer}, {Hu}, {Langer}, {Townsend},
  {Townsley}, \& {Timmes}}]{Paxton15}
{Paxton}, B., {Marchant}, P., {Schwab}, J., {et~al.} 2015, ApJS, 220, 15,
  \dodoi{10.1088/0067-0049/220/1/15}

\bibitem[{{Paxton} {et~al.}(2018){Paxton}, {Schwab}, {Bauer}, {Bildsten},
  {Blinnikov}, {Duffell}, {Farmer}, {Goldberg}, {Marchant}, {Sorokina},
  {Thoul}, {Townsend}, \& {Timmes}}]{Paxton18}
{Paxton}, B., {Schwab}, J., {Bauer}, E.~B., {et~al.} 2018, ApJS, 234, 34,
  \dodoi{10.3847/1538-4365/aaa5a8}

\bibitem[{{Paxton} {et~al.}(2019){Paxton}, {Smolec}, {Schwab}, {Gautschy},
  {Bildsten}, {Cantiello}, {Dotter}, {Farmer}, {Goldberg}, {Jermyn}, {Kanbur},
  {Marchant}, {Thoul}, {Townsend}, {Wolf}, {Zhang}, \& {Timmes}}]{Paxton19}
{Paxton}, B., {Smolec}, R., {Schwab}, J., {et~al.} 2019, ApJS, 243, 10,
  \dodoi{10.3847/1538-4365/ab2241}

\bibitem[{{Pietrukowicz} {et~al.}(2017){Pietrukowicz}, {Dziembowski}, {Latour},
  {Angeloni}, {Poleski}, {di Mille}, {Soszy{\'n}ski}, {Udalski},
  {Szyma{\'n}ski}, {Wyrzykowski}, {Koz{\l}owski}, {Skowron}, {Skowron},
  {Mr{\'o}z}, {Pawlak}, \& {Ulaczyk}}]{Pietrukowicz2017}
{Pietrukowicz}, P., {Dziembowski}, W.~A., {Latour}, M., {et~al.} 2017, Nature
  Astronomy, 1, 0166, \dodoi{10.1038/s41550-017-0166}

\bibitem[{{Pigulski} {et~al.}(2022){Pigulski}, {Kotysz}, \&
  {Ko{\l}aczek-Szyma{\'n}ski}}]{Pigulski2022}
{Pigulski}, A., {Kotysz}, K., \& {Ko{\l}aczek-Szyma{\'n}ski}, P.~A. 2022, A\&A,
  663, A62, \dodoi{10.1051/0004-6361/202243293}

\bibitem[{{Preece} {et~al.}(2022){Preece}, {Hamers}, {Battich}, \&
  {Rajamuthukumar}}]{Preece2022}
{Preece}, H.~P., {Hamers}, A.~S., {Battich}, T., \& {Rajamuthukumar}, A.~S.
  2022, MNRAS, 517, 2111, \dodoi{10.1093/mnras/stac2798}

\bibitem[{{Ramsay} {et~al.}(2022){Ramsay}, {Woudt}, {Kupfer}, {van Roestel},
  {Paterson}, {Warner}, {Buckley}, {Groot}, {Heber}, {Irrgang}, {Jeffery},
  {Motsoaledi}, {Schwartz}, \& {Wevers}}]{Ramsay2022}
{Ramsay}, G., {Woudt}, P.~A., {Kupfer}, T., {et~al.} 2022, MNRAS, 513, 2215,
  \dodoi{10.1093/mnras/stac1000}

\bibitem[{{Romero} {et~al.}(2018){Romero}, {C{\'o}rsico}, {Althaus},
  {Pelisoli}, \& {Kepler}}]{Romero2018}
{Romero}, A.~D., {C{\'o}rsico}, A.~H., {Althaus}, L.~G., {Pelisoli}, I., \&
  {Kepler}, S.~O. 2018, MNRAS, 477, L30, \dodoi{10.1093/mnrasl/sly051}

\bibitem[{{Saio} \& {Jeffery}(2000)}]{Saio2000}
{Saio}, H., \& {Jeffery}, C.~S. 2000, MNRAS, 313, 671,
  \dodoi{10.1046/j.1365-8711.2000.03221.x}

\bibitem[{{Shen} {et~al.}(2009){Shen}, {Idan}, \& {Bildsten}}]{Shen2009}
{Shen}, K.~J., {Idan}, I., \& {Bildsten}, L. 2009, ApJ, 705, 693,
  \dodoi{10.1088/0004-637X/705/1/693}

\bibitem[{{Sweigart}(1997)}]{Sweigart1997}
{Sweigart}, A.~V. 1997, ApJL, 474, L23, \dodoi{10.1086/310414}

\bibitem[{{Townsend} \& {Teitler}(2013)}]{Townsend2013}
{Townsend}, R.~H.~D., \& {Teitler}, S.~A. 2013, MNRAS, 435, 3406,
  \dodoi{10.1093/mnras/stt1533}

\bibitem[{{Wu} \& {Li}(2018)}]{Wu2018}
{Wu}, T., \& {Li}, Y. 2018, MNRAS, 478, 3871, \dodoi{10.1093/mnras/sty1347}

\bibitem[{{Xiong} {et~al.}(2022){Xiong}, {Casagrande}, {Chen}, {Vos}, {Zhang},
  {Justham}, {Li}, {Wu}, {Li}, \& {Han}}]{Xiong2022}
{Xiong}, H., {Casagrande}, L., {Chen}, X., {et~al.} 2022, arXiv e-prints,
  arXiv:2211.01564.
\newblock \doarXiv{2211.01564}

\bibitem[{{Zhang} {et~al.}(2017){Zhang}, {Hall}, {Jeffery}, \&
  {Bi}}]{Zhang2017}
{Zhang}, X., {Hall}, P.~D., {Jeffery}, C.~S., \& {Bi}, S. 2017, ApJ, 835, 242,
  \dodoi{10.3847/1538-4357/835/2/242}

\bibitem[{{Zhang} \& {Jeffery}(2012)}]{Zhang12}
{Zhang}, X., \& {Jeffery}, C.~S. 2012, MNRAS, 419, 452,
  \dodoi{10.1111/j.1365-2966.2011.19711.x}

\bibitem[{{Zorotovic} {et~al.}(2011){Zorotovic}, {Schreiber}, \&
  {G{\"a}nsicke}}]{Zorotovic2011}
{Zorotovic}, M., {Schreiber}, M.~R., \& {G{\"a}nsicke}, B.~T. 2011, A\&A, 536,
  A42, \dodoi{10.1051/0004-6361/201116626}

\end{thebibliography}

\end{document}